\documentclass[thmsa,12pt]{article}
\usepackage{sw20lart}

\input{tcilatex}
\begin{document}

\author{Yu.V.Kozlov,S.V.Khaltourtcev,I.N.Machulin,A.V.Martemyanov, \and %
V.P.Martemyanov,S.V.Sukhotin,V.G.Tarasenkov,E.V.Turbin, \and V.N.Vyrodov.
\and Russian Research Center ''Kurchatov Institute'' \and 123182 Moscow,
Russia.}
\title{ANTINEUTRINO-DEUTERON EXPERIMENT AT KRASNOYRSK REACTOR.}
\date{1999 year.}
\maketitle

\begin{abstract}
The investigation of antineutrino interaction with matter at Krasnoyarsk
reactor is described. The characteristics of the detector ''Deuteron'' and
the present results and perspectives are discussed.
\end{abstract}

\section{INTRODUCTION.}

This report is represented the results of some experiments, which carried
out at the neutrino underground laboratory of Kranoyarsk nuclear plant.

At the first it is necessary to say about the specific condition on
Krasnoyarsk reactors:

\begin{itemize}
\item  the unique complex of the industrial nuclear reactors is inside rock
and a passive shielding from cosmic muons corresponds to 600 m.w.e. Due to
the muons flux is suppressed by a factor 1000;

\item  the composition of a nuclear fuel in the reactor is such, that the
difference between the real antineutrino spectrum and U-235 spectrum is less
then 1\%;

\item  the peroid ''reactor-on'' is equal to approximately 50 days and
therefore one may measure the background each two months but not 1-1.5 years
usual on power atomic station.
\end{itemize}

The new experiment for studding of interaction antineutrino with a deuteron
on the improvement detector ''Deuteron'' is in progress now.

The interaction of antineutrinos ($\stackrel{\symbol{126}}{\upsilon _{e}}$ )
with a deuteron occurs via two channels, Neutral Current on Deuteron (NCD)
and Charged Current on Deuteron (CCD),

\begin{center}
$\stackrel{\symbol{126}}{v_{e}}+\ d\rightarrow p+n+\stackrel{\symbol{126}}{v}%
_{e}^{^{\prime }}\qquad \left( 1\right) $

$\stackrel{\symbol{126}}{v_{e}}+\ d\rightarrow n+n+e^{+}\qquad \left(
2\right) $
\end{center}

The study of these reactions can give the information about:

\begin{itemize}
\item  a) weak constants for charged and neutral currents;

\item  b) a length of neutron-neutron scattering;

\item  c) neutrino oscillation.
\end{itemize}

The results of the previous experiments on the study of
antineutrino-deuteron interaction are shown in Table 1.

\begin{center}
Table 1.

\begin{tabular}{|l|l|l|}
\hline
Savannah River \cite{1} & $\sigma ^{ncd}=3.8\pm 0.9$ & $\sigma _{\exp
}^{ncd}/\sigma _{theor}^{ncd}=0.8\pm 0.2$ \\ \cline{2-3}
$\sigma [10^{-45}cm^{2}/v_{e}]$ & $\sigma ^{ccd}=1.5\pm 0.4$ & $\sigma
_{\exp }^{ccd}/\sigma _{theor}^{ccd}=0.7\pm 0.2$ \\ \cline{2-3}
& $\sigma _{\exp }^{ccd}/\sigma _{\exp }^{ncd}=0.40\pm 0.14$ & $\sigma
_{thoer}^{ccd}/\sigma _{theor}^{ncd}=0.353$ \\ \hline
Krasnoyarsk \cite{2} & $\sigma ^{ncd}=3.0\pm 1.0$ & $\sigma _{\exp
}^{ncd}/\sigma _{theor}^{ncd}=0.95\pm 0.33$ \\ \cline{2-3}
$\sigma [10^{-44}cm^{2}/fis.^{235}U]$ & $\sigma ^{ccd}=1.1\pm 0.2$ & $\sigma
_{\exp }^{ccd}/\sigma _{theor}^{ccd}=0.98\pm 0.18$ \\ \cline{2-3}
& $\sigma _{\exp }^{ccd}/\sigma _{\exp }^{ncd}=0.37\pm 0.14$ & $\sigma
_{thoer}^{ccd}/\sigma _{theor}^{ncd}=0.353$ \\ \hline
Rovno \cite{3} & $\sigma ^{ncd}=2.71\pm 0.46\pm 0.11$ & $\sigma _{\exp
}^{ncd}/\sigma _{theor}^{ncd}=0.92\pm 0.18$ \\ \cline{2-3}
$\sigma [10^{-44}cm^{2}/PWR$-$440]$ & $\sigma ^{ccd}=1.17\pm 0.14\pm 0.07$ & 
$\sigma _{\exp }^{ccd}/\sigma _{theor}^{ccd}=1.08\pm 0.19$ \\ \cline{2-3}
& $\sigma _{\exp }^{ccd}/\sigma _{\exp }^{ncd}=0.43\pm 0.10$ & $\sigma
_{thoer}^{ccd}/\sigma _{theor}^{ncd}=0.37\pm 0.08$ \\ \hline
Bugey \cite{4} & $\sigma ^{ncd}=3.29\pm 0.42$ & $\sigma _{\exp
}^{ncd}/\sigma _{theor}^{ncd}=1.01\pm 0.13$ \\ \cline{2-3}
$\sigma [10^{-44}cm^{2}/fis.]$ & $\sigma ^{ccd}=1.10\pm 0.23$ & $\sigma
_{\exp }^{ccd}/\sigma _{theor}^{ccd}=0.97\pm 0.20$ \\ \cline{2-3}
& $\sigma _{\exp }^{ccd}/\sigma _{\exp }^{ncd}=0.33\pm 0.08$ & $\sigma
_{thoer}^{ccd}/\sigma _{theor}^{ncd}=0.348\pm 004$ \\ \hline
\end{tabular}
\end{center}

\section{DETECTOR DESIGN.}

The modernized detector ``Deuteron''(Fig.1) is situated in the underground
laboratory at a distance 34.0 m from the reactor, the neutrino flux is about
a few units to 10$^{12}$ $\stackrel{\symbol{126}}{v}_{e}/cm^{2}.$

The target volume is 513 l of D2O (H2O) placed in a stainless tank, which is
surrounded by 30 cm of Teflon for neutron reflections, 0.1 cm of Cd, 8 cm of
steel shots, 20 cm of graphite and 16 cm of boron polyethylene (CH2+3\%B)
for gamma and neutron shielding. The whole installation is pierced to make
169 holes (81 holes pass through the Tank and Teflon, the others through the
Teflon only). These holes house 169 proportional 3He neutron counters with a
reduced intrinsic alpha background. These counters are used for neutron
registrations. They are located in a square lattice with a side of 10 cm.
The active shielding covers the main assembly, against cosmic muons.

The neutron counters used in theexperiment canregister only neutrons, so
this detector is a detector of a integral type. The counter consisits of a
stainless steel tube 1 m long and 31 mm in diameter with walls of 0.5 mm
thick. A 20-$\mu m$ wire is strectched along the counter. The wire is made
from tungsten and coated with gold. The inner surface of counter is covered
with 60-$\mu m$ Teflon layer to reduce the natural alpha-background from the
stainless steel wall, and than the Teflon layer is covered with 2 $\mu m$
pure copper layer to keep the counter able to work.The counter is filled
with a mixture of 4 KPa $^{3}$He and 4 KPa $^{40}$Ar gases.

\section{THE CHARACTERISTICS OF THE DETECTOR AND MONTE-CARLO SIMULATIONS.}

A reaction is used for detecting neutrons. An amplitude spectrum is shown on
the Fig.2. The ''wall'' effect or losses of a part of energy in the counter
wall have been measured and is shown on the Fig.3

An efficiency of the detector was calculated with Monte-Carlo method as for
inverse beta decay reaction as for antineutrino-deuteron reaction. Also
calculations

have been made for Cf252 source and this result was checked experimentally.
The difference (less 1\%) between the calculation and experimental data
shows good reliability of MC calculations. The neutron efficiency and
neutron livetime are shown in the Table 2.

\begin{center}
Table 2.

\begin{tabular}{|p{6cm}|l|l|}
\hline
Parameters/Target & H$_{2}$O & D$_{2}$O \\ \hline
Efficiency of one neutron registration by tank counters only & $(27.5\pm
0.3)\%$ & $\left( 56.0\pm 0.7\right) \%$ \\ \hline
Efficiency of double neutron registration by all counters & $\left( 9.9\pm
0.1\right) \%$ & $\left( 41.6\pm 0.4\right) \%$ \\ \hline
Neutron livetime & $\left( 138\pm 2\right) \,\mu s$ & $\left( 203\pm
2\right) \,\mu s$ \\ \hline
\end{tabular}
\end{center}

A special attention was given to the correlated background for NCD channel
connected with the antineutrino interaction with proton (H$_{2}$ atoms),
because the cross section for such process is relatively large. The
construction of the detector allowed us to decrease the efficiency neutron
registration from a boron polyethylene up to 0.002\%. As a result we
estimate the correlated background (Ncor) as 0.6 events/day due to the
concentration light water in heavy water is 0.15\%.

\section{EXPERIMENT.}

\subsection{Data collection system.}

The experiment was monitored ON-LINE on the CAMAC.

The event is registration of a neutron in the detector. The total
information about event include itself:

\begin{itemize}
\item  amplitude of any neutron;

\item  astronomic time;

\item  neutron zone registration (detector was divided into 32 groups of
counters);

\item  multiplicity of event (a number of neutrons in 800 ms registration
window after the first occurred neutron in the event);

\item  condition of event (no veto comes in the 800 ms interval before and
after any neutron);

\item  time between neutrons in the same event.
\end{itemize}

\subsection{Target H$_{2}$O.}

The inverse beta decay on the proton reaction

\begin{center}
$\stackrel{\symbol{126}}{v_{e}}+\ p\ \longrightarrow \ n\ +\ e^{+}\qquad (3)$%
\mathstrut
\end{center}

is used for checking and improving of some parameters of the detector. The
exposure time is 115$\times 10^{-45}$s. or about 133 days. There are 4 sets
of measurement with different background condition have been made. The
results are represented in the Table 3.

\begin{center}
Table 3.

\begin{tabular}{|l|l|l|l|}
\hline
SET & \multicolumn{2}{|l|}{Reactor power} & Effect \\ \cline{2-3}
& ON & OFF &  \\ \hline
I & $403.5\pm 4.5$ & $201.4\pm 7.5$ & $202.1\pm 8.0$ \\ \hline
II & $395.5\pm 3.6$ & $204.9\pm 7.5$ & $190.7\pm 7.7$ \\ \hline
III & $381.4\pm 3.9$ & $187.9\pm 7.5$ & $196.2\pm 6.9$ \\ \hline
IV & $379.0\pm 4.6$ & $169.6\pm 5.6$ & $209.4\pm 7.3$ \\ \hline
$\sum $ & \multicolumn{2}{|l|}{} & $205.1\pm 3.8$ \\ \hline
\end{tabular}
\end{center}

The results was obtained with following cuts:

\begin{itemize}
\item  only tank events were analyzed;

\item  the amplitude region of neutron registration with 644 to 884 KeV was
taken.
\end{itemize}

In result the CCP cross section is

\begin{center}
$\sigma _{\exp }^{ccp}\ =\ \left( 6.39\times 10^{-43}\pm 3.0\%\right) \
cm^{2}/fission\ ^{235}U$
\end{center}

This result is in a good agreement with theoretical cross section (V-A
theory). The ratio is (68\% C.L.):

\begin{center}
$R\ =\ \frac{\sigma _{\exp }^{ccp}}{\sigma _{v-a}}\left( ^{235}U\right) \ =\
1.00\ \pm 0.04$
\end{center}

\subsection{Target D$_{2}$O.}

From the beginning of 1997 and up to now the antineutrino-deuteron
experiment is in progress. Data have been collected during 360 days reactor
''on'' and 120 days reactor ''off''. There were 8 sets of measurements. The
results are shown in the Table 4.

\begin{center}
Table 4.

\begin{tabular}{|c|c|c|c|c|}
\hline
SET & \multicolumn{2}{|c}{T measured, 10$^{5}\sec $} & \multicolumn{2}{|c|}{
Effect per $10^{5}\sec $} \\ \cline{2-5}
& Reactror ''ON'' & Reactor ''OFF'' & NCD (only tank) & CCD \\ \hline
I & 27.96 & 13.77 & $23.25\pm 5.97$ & $3.37\pm 1.46$ \\ \hline
II & 34.94 & 10.16 & $21.14\pm 6.38$ & $3.93\pm 1.55$ \\ \hline
III & 26.82 & 5.94 & $11.79\pm 8.19$ & $4.16\pm 1.87$ \\ \hline
IV & 45.04 & 20.44 & $28.38\pm 5.72$ & $4.27\pm 1.15$ \\ \hline
V & 59.43 & 8.75 & $22.67\pm 7.76$ & $4.14\pm 1.53$ \\ \hline
VI & 62.10 & 24.10 & $28.15\pm 5.10$ & $4.91\pm 1.00$ \\ \hline
VII & 28.26 & 18.94 & $17.33\pm 6.37$ & $5.03\pm 1.20$ \\ \hline
VIII & 43.34 & 9.90 & $26.15\pm 7.17$ & $4.75\pm 1.37$ \\ \hline
$\sum $ & 327.89 & 112.00 & $24.39\pm 2.24$ & $4.44\pm 0.47$ \\ \hline
\end{tabular}
\end{center}

We used next cuts:

\begin{itemize}
\item  evnts with the amplitude of the first neutron in interval from 644 to
844 KeV and the amplitude of second neutron in interval from 190 to 884 Kev
and with the time between of two neutrons in region from 5 to 800 $\mu \sec $
selected for CCD channel;

\item  Events are detected by tank of the detector selected for NCD reaction.
\end{itemize}

To be sure that electronic and background conditions are quite stable the
analyses of events with neutron multiplicity 3 and more was performed.
Results are shown in Table 5.

\begin{center}
Table 5.

\begin{tabular}{|c|c|c|c|}
\hline
Multiplicity & ''ON'' & ''OFF'' & ''ON''-''OFF'' \\ \hline
3 & $3.108\pm 0.097$ & $3.115\pm 0.169$ & $-0.007\pm 0.19$ \\ \hline
more then 2 & $4.70\pm 1.12$ & $4.68\pm 0.21$ & $0.02\pm 0.24$ \\ \hline
\end{tabular}
\end{center}

\subsection{Primary results.}

Take to account both ''wall'' effect and time rejection for double neutron
events and amplitude selection efficiencies registration of neutron are:

\begin{center}
$\varepsilon _{1}^{ncd}\ $(all detector) $=\ 0.584$

$\varepsilon _{1}^{ccd}\ $(all detector) $=\ 0.584\ $(registration the first
neutron)

$\varepsilon _{2}^{ccd}\ $(all detector) $=\ 0.619\ $(registration the
second neutron)

$\varepsilon _{1}^{ncd}\ $(tank) $=\ 0.507$
\end{center}

After correction on probability registration in NCD channel events
corresponded CCD channel and correlated background receive that

\begin{center}
$N^{ncd}(NCD)=18.30\pm 1.71\qquad \left( 4\right) \ \;$(only tank)
\end{center}

From following below formulas (5), (6) and (7)

\begin{center}
$N^{ncd}=P_{reactor}\times \varepsilon _{1}^{ncd}\times N_{d}\times \sigma
_{\exp }^{ncd}\ \left( 5\right) $

$N^{ccd}=P_{reactor}\times \varepsilon _{1}^{ccd}\times N_{d}\times \sigma
_{\exp }^{ccd}\ \left( 6\right) $

$N^{ccp}=P_{reactor}\times \varepsilon _{1}^{ccp}\times N_{p}\times \sigma
_{\exp }^{ccd}\ \left( 7\right) $
\end{center}

can obtain that

\begin{center}
$\sigma _{\exp }^{ncd}=(3.09\pm 0.30)\times 10^{-44}\ cm^{2}\ /\ fission\
^{235}U$

$\sigma _{\exp }^{ccd}=(1.05\pm 0.12)\times 10^{-44}\ cm^{2}\ /\ fission\
^{235}U$
\end{center}

These results are in good agreement with theory.

\begin{center}
Table 6.

\begin{tabular}{|c|c|c|}
\hline
& \multicolumn{2}{|c|}{$\sigma ,\times 10^{-44}\ cm^{2}\ /\ fission\ ^{235}U$%
} \\ \cline{2-3}
& NCD & CCD \\ \hline
Experiment & $3.09\pm 0.30$ & $1.05\pm 0.12$ \\ \hline
Theory (Schreckhenbach spectrum) & $3.18\pm 0.17$ & $1.07\pm 0.07$ \\ \hline
Ratio (experiment/theory) & $0.97\pm 0.11$ & $0.99\pm 0.13$ \\ \hline
\end{tabular}
\end{center}

\section{FUTURE.}

We plan to continue experiment up to 2000 year, so about 500 days and 170
days reactor ''on'' and ''off'' respectively will be taken. It gives us the
decreasing of error up to 8\% for NCD and CCD channels.

To decrease the statistic error for CCD channel we plane to reject events
using the geometry of detected neutrons.

After usual measurements will be made the calibration runs with Cf source
installed in each hall to check Monte-Carlo simulation.

\section{ACKNOWLEGEMENTS.}

This work is supported by grants RFFI 99-15-96640 and 98-02-16313.

We would like to thank the staff of Krasnoyarsk reactor for constant help,
Ac. S.T. Belayev and Dr. Yu.V.Gaponov for very useful discusions.

\begin{tabular}{|c|c|c|}
\hline
& \multicolumn{2}{|c|}{$\sigma ,\times 10^{-44}\ cm^{2}\ /\ fission\ ^{235}U$%
} \\ \cline{2-3}
& NCD & CCD \\ \hline
Experiment & $3.09\pm 0.30$ & $1.05\pm 0.12^{*}$ \\ \hline
Theory (Schreckhenbach spectrum) & $3.18\pm 0.17$ & $1.07\pm 0.07$ \\ \hline
Ratio (experiment/theory) & $0.97\pm 0.11$ & $0.99\pm 0.13$ \\ \hline
\end{tabular}

\end{document}